\documentclass[useAMS,usegraphicx]{mn2e}
\usepackage{aas_macros}
\usepackage{url}
\def\etal{{et~al. }}
\def\OII{{[O~\sc{ii}]}}
\def\hdel{{H${\delta}$}}
\def\EWH{{EW\{\hdel\}}}
\def\EWO{{EW\{\OII\}}}
\def\numl{{5,697}}

\title[The Star Formation Histories of Luminous Red Galaxies]{The 2dF-SDSS LRG and QSO Survey: The Star Formation Histories of Luminous Red Galaxies}
\author[Roseboom et al.] 
{Isaac G. Roseboom$^{1}$\thanks{E-mail:roseboom@physics.uq.edu.au}, Kevin A. Pimbblet$^{1}$, Michael J. Drinkwater$^{1}$, Russell D. Cannon$^{2}$,\newauthor Roberto De Propris$^{3}$, Alastair C. Edge$^{4}$, Daniel J. Eisenstein$^{5}$, Robert C. Nichol$^{6}$, \newauthor Ian Smail$^{4}$, David A. Wake$^{7}$, Joss Bland Hawthorn$^{2}$, Terry J. Bridges$^{8}$, \newauthor Daniel Carson$^{6}$, Matthew Colless$^{2}$, Warrick J. Couch$^{9}$, Scott M. Croom$^{2}$, \newauthor Simon P. Driver$^{10}$, Paul C. Hewett$^{11}$, Jon Loveday$^{12}$, Nic Ross$^{7}$,\newauthor Donald P. Schneider$^{13}$, Tom Shanks$^{7}$, Robert G. Sharp$^{2}$, Peter Weilbacher$^{7,14}$\\
$^{1}$Department of Physics, University of Queensland, QLD 4072, Australia\\
$^{2}$Anglo-Australian Observatory, PO Box 296, Epping, NSW 1710, Australia\\
$^{3}$Cerro Tololo Inter-American Observatory, Casilla 603, La Serena, Chile\\
$^{4}$Institute for Computational Cosmology, Durham University, South Road, Durham DH1 3LE\\
$^{5}$Steward Observatory, 933 N. Cherry Ave, Tucson, AZ 85721, USA\\
$^{6}$Institute of Cosmology and Gravitation, University of Portsmouth, PO1 2EG \\
$^{7}$Department of Physics, Durham University, South Road, Durham DH1 3LE\\
$^{8}$Physics Department, Queen's University, Kingston, ON, K7L 3N6, Canada\\
$^{9}$Department of Astrophysics, University of New South Wales, Sydney, NSW 2052, Australia\\
$^{10}$Research School of Astronomy and Astrophysics, Australian National University, Canberra, ACT 2600, Australia\\
$^{11}$Institute of Astronomy, Madingley Road, Cambridge CB3 0HA\\
$^{12}$Dept of Physics \& Astronomy, University of Sussex, Falmer, Brighton BN1 9QH\\
$^{13}$Department of Astronomy and Astrophysics, The Pennsylvania State University, University Park, PA 16802, USA\\
$^{14}$Astrophysikalisches Institut Potsdam, An der Sternwarte 16, D-14482 Potsdam, Germany\\  
}
\begin{document}

\date{\today}

\pagerange{\pageref{firstpage}--\pageref{lastpage}} \pubyear{2006}

\maketitle

\label{firstpage}
\begin{abstract}
%FIX!!!!!!!!
We present a detailed investigation into the recent star formation histories of \numl\/ Luminous Red Galaxies (LRGs) based on the \hdel (4101\AA), and \OII\/ (3727\AA) lines and the D4000 index. LRGs are luminous (L$>3$L*), galaxies which have been selected to have photometric properties consistent with an old, passively evolving stellar population. For this study we utilise LRGs from the recently completed 2dF-SDSS LRG and QSO survey (2SLAQ). Equivalent widths of the \hdel\/ and \OII\/ lines are measured and used to define three spectral types, those with only strong \hdel\/ absorption (k+a), those with strong \OII\/ in emission (em) and those with both (em+a). All other LRGs are considered to have passive star formation histories. The vast majority of LRGs are found to be passive ($\sim$80 per cent), however significant numbers of k+a (2.7 per cent), em+a (1.2 per cent) and em LRGs (8.6 per cent) are identified. An investigation into the redshift dependence of the fractions is also performed. A sample of SDSS MAIN galaxies with colours and luminosities consistent with the 2SLAQ LRGs is selected to provide a low redshift comparison. While the em and em+a fractions are consistent with the low redshift SDSS sample, the fraction of k+a LRGs is found to increase significantly with redshift. This result is interpreted as an indication of an increasing amount of recent star formation activity in LRGs with redshift. By considering the expected life time of the k+a phase, the number of LRGs which will undergo a k+a phase can be estimated. A crude comparison of this estimate with the predictions from semi-analytic models of galaxy formation shows that the predicted level of k+a and em+a activity is not sufficient to reconcile the predicted mass growth for massive early-types in a hierarchical merging scenario.

\end{abstract}

\begin{keywords}
galaxies: evolution -- galaxies: formation -- surveys
\end{keywords}

\section{Introduction}  

The widely accepted paradigm for galaxy formation is the hierarchical merger model which dictates that galaxies form via continuous mergers with smaller objects (e.g. \nocite{cole1994}{Cole} {et~al.} 1994). This scenario is a natural consequence of the $\Lambda$ cold dark matter ($\Lambda$CDM) model of structure formation, which has been highly successful in explaining both the observed structure from the large redshift surveys (2dF Galaxy Redshift Survey (2dFGRS); \nocite{Colless01}{Colless} {et~al.} 2001, Sloan Digital Sky Survey (SDSS); \nocite{stoughton2002}{Stoughton} {et~al.} 2002), and the anisotropy in the cosmic microwave background \nocite{Sper03}({Spergel} {et~al.} 2003). In the context of this model one would naively expect the most massive galaxies to also be the youngest as they have been `built up' slowly via mergers.
However traditionally early-types are thought to have formed early and rapidly, with a vast amount of observational evidence points supporting this conclusion. The homogeneous nature of massive early-type galaxy properties, such as small scatter in the colour-magnitude relation (CMR) \nocite{Vis77,Bower92,Terl2001,aragon93,stanford95,ellis97,stanford98}({Visvanathan} \& {Sandage} 1977; {Bower}, {Lucey} \& {Ellis} 1992; {Terlevich}, {Caldwell}, \&  {Bower} 2001; {Aragon-Salamanca} {et~al.} 1993; {Stanford}, {Eisenhardt}, \&  {Dickinson} 1995; {Ellis} {et~al.} 1997; {Stanford}, {Eisenhardt}, \&  {Dickinson} 1998, among others) and fundamental plane (FP) \nocite{Dress1987,Burstein1997}({Dressler} {et~al.} 1987; {Burstein} {et~al.} 1997), and the lack of evolution in these relations with redshift \nocite{Kodama1998,Vandokkum2003}({Kodama} {et~al.} 1998; {van Dokkum} \& {Stanford} 2003), is consistent with a scenario in which most, if not all massive early-types, formed at high redshift (at least $z>2$) and have experienced only passive evolution of their stellar populations since.

While this could simply be a result of selection bias \nocite{dokkum2001}({van Dokkum} \& {Franx} 2001), the lack of direct evidence of strong evolution in the population of early-type galaxies since $z\sim$1 remains a serious problem for the hierarchical model of galaxy formation. In recent years much work has gone into reconciling these differences. While theoretical models have moved towards consistency with the CMR and FP \nocite{Kauffmann98, delucia2005}({Kauffmann} \& {Charlot} 1998; {De Lucia} {et~al.} 2005, among others), observational studies based on the spectroscopic and morphological properties of early-types have found significant populations which show evidence for recent star formation and/or interactions. A morphologically based study by \nocite{michard2004}{Michard} \& {Prugniel} (2004) found a significant number ($\sim30$ per cent) of nearby ($z\sim0.01$) early-type galaxies showed evidence of recent interactions/mergers, although 40 per cent of these showed no evidence for a young stellar population. A similar study of 86 nearby bulge-dominated red galaxies by \nocite{vandokkum2005} {van Dokkum} {et~al.} (2005) found 71 per cent had evidence of tidal interactions. However given that both of these studies were undertaken at low redshift the volume sampled was too small for significant numbers of LRG analogues to be present. Moving to higher redshifts, \nocite{Willis2002}{Willis} {et~al.} (2002) found evidence for current star formation, in the form of detection of the \OII\/ emission line, in over 20 per cent of a sample of 415 luminous field early-types at $z\sim0.3$. While these studies show that evolution is occurring in at least some of the early-type population, it remains unclear if all early-types have experienced evolution of this type. Another contentious point is whether early-type galaxy acquire mass predominately in the form of star forming gas or already formed stars via so called `dry mergers' (\nocite{bell2005} Bell  et al. 2005).

In this study we look for evidence of recent star formation in a sample of Luminous Red Galaxies (LRGs). LRGs are luminous ($L>3L^*$) early-type galaxies, analogous to bright cluster galaxies (BCG), selected via their red colours which are broadly consistent with a passively evolving stellar population. These galaxies are ideal candidates for the purposes of this study as they have been shown to have relatively homogeneous spectral properties \nocite{eis01} ({Eisenstein} {et~al.} 2001), enabling any deviation due to recent star formation episodes to be readily identified. They are also the most massive galaxies, which, if the hierarchical merger model is to be correct, must also have the most vigorous merger histories. Thus a sufficiently large sample of LRGs provides the perfect test bed for current theories of galaxy formation/evolution.

Here we probe the star formation histories of our LRG sample via their spectral properties, in particular the \OII\/(3727\AA) emission and \hdel\/(4101\AA) absorption lines and the D4000 index. LRGs are selected from both the SDSS and 2dF-SDSS LRG and QSO (2SLAQ) surveys, resulting in a sample of \numl\/ LRGs covering a redshift range of $0.2<z<0.7$. Details of the sample selection are presented in Section \ref{sec:sam}. Section \ref{sec:meth} describes the analysis and modeling performed, with the subsequent results presented in Section \ref{sec:results}. The implications and conclusions to be drawn from these results are then presented in Sections \ref{sec:disc}, and \ref{sec:conc}, respectively. 

Where necessary, we have assumed a flat cosmology with $\Omega_m=0.3$, $\Omega_{\Lambda}=0.7$ and H$_0=70$ kms$^{-1}$Mpc$^{-1}$.

\section{The 2SLAQ LRG survey} \label{sec:sam}

Completed in August 2005, the 2SLAQ survey has collected spectra and measured redshifts for over 10,000 high redshift QSOs and $\sim$ 14,000 LRGs in the redshift range $0.45<z<0.8$. The survey is essentially a high redshift extension to the SDSS LRG survey, utilising the 2dF facility on the Anglo-Australian Telescope to go deeper than was possible with the SDSS survey telescope. The 2SLAQ dataset has already been utilised to produce photometric redshifts (\nocite{pad04}{Padmanabhan} {et~al.} 2005), and a large photo-z LRG catalogue (Lahav \etal\/ 2006), as well as probe the evolution of the LRG luminosity function from $z=0.2$ to $z=0.55$ (Wake \etal\/ 2006). 

\begin{figure}
%\centering
%\begin{minipage}{140mm}
  \includegraphics[angle=270,scale=0.6]{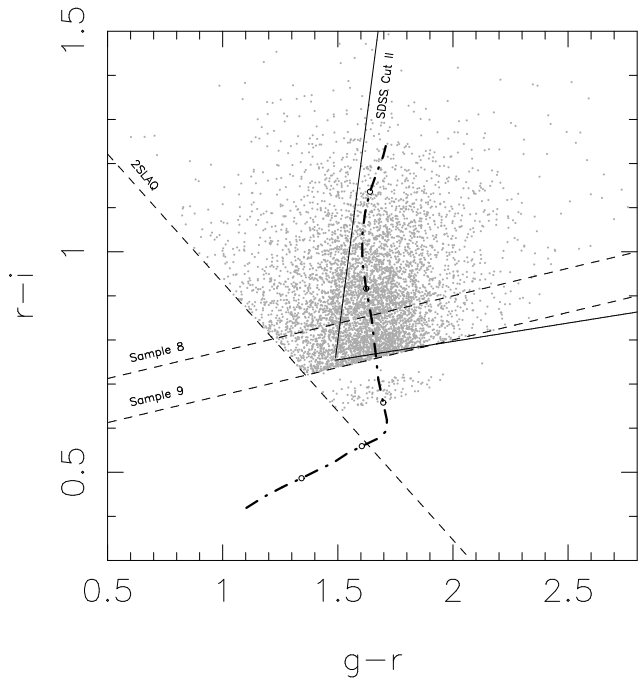}
  \caption{Colour-colour diagram for LRGs selected in this study. The solid lines show the region selected by the SDSS cut II, the dashed lines represent the 2SLAQ selections. The dot-dashed line shows the track taken by a passively evolving early-type galaxy model. The circles are located at $\Delta z=0.1$ intervals, starting at $z=0.2$ (lower left) and ending at $z=0.6$ (top right). The sharp bend in the track near $z=0.4$ occurs where the 4000\AA\/ break goes from the $g$ to the $r$ band. }
\label{fig:ccpaper}
%\end{minipage}
\end{figure}

While a detailed description of the survey is presented in Cannon \etal (2006), we briefly review the pertinent points here. The 2SLAQ survey was designed to be an extension of the SDSS LRG survey which sampled LRGs in the redshift range $0.15<z<0.5$, to higher redshifts (\nocite{eis01}{Eisenstein} {et~al.} 2001). The SDSS LRG survey utilised two colour selections in $g-r$ and $r-i$ colour space. One colour selection was used to sample LRGS at redshifts less than 0.4 (cut I), where the 4000\AA\/ break features prominently in the $g$ band, while another (cut II) was used for higher redshifts, where the 4000\AA\/ break is located in the $r$ band. The two cuts were necessary as the 4000\AA\/ break dominates the colour evolution of massive early-type galaxies, such as LRGs, and as such the transition from one band to another causes the track taken by the galaxy in colour-colour space to change considerably. In addition to the colour cut, a sliding magnitude limit was utilised for the low redshift cut I LRGs as the colours of LRGs at $z<0.4$ are quite similar to less luminous star forming galaxies at lower redshifts. The 2SLAQ survey utilised a colour selection similar to the higher redshift, cut II, SDSS colour cut with some key differences. As the number of fibres available to 2SLAQ survey per sq deg. was significantly different ($\sim 67$ per sq deg. for 2SLAQ as compared to only $\sim 2$ per sq deg. for SDSS cut II targets) the colour selection has to be `loosened' to provide enough targets to fill a 2dF field. This can be seen in Figure \ref{fig:ccpaper} which shows the position of the 2SLAQ LRGs used in this study with respect to both the 2SLAQ and SDSS cut II colour cuts. A passive evolution track, provided by Bruzual and Charlot (2003) models, is shown as the solid dashed line. The SDSS LRG survey, with its smaller number of targets per field cuts very close to the passive evolution track and thus goes close to only selecting `red and dead', passively evolving, galaxies. By contrast the increased availability of fibres from 2dF allows the 2SLAQ selection to be quite broad in $g-r$ colour. This has a distinct advantage over SDSS in that LRGs with a range of star formation histories are allowed by the 2SLAQ selection, not just strictly passive evolution. In addition to the colour cuts a magnitude limit of $i_{Dev}<19.8$ is imposed so as to ensure that reasonable signal to noise ($\sim3$) could be achieved in 4 hrs.  

Another key difference between the 2SLAQ and SDSS cut II selections is the grouping and prioritization of targets based on a number of colour cuts. LRGs which lie above the line marked 'Sample 8' were the main goal of the 2SLAQ survey and given the highest priority in targeting. The jump in $r-i$ from the SDSS cut II selection and the 2SLAQ sample 8 cut is a result of the desire to acquire distinctly higher redshift LRGs than those in SDSS. Sample 9 LRGs were given lower priority and are essentially lower redshift targets closer to the redshift range of the SDSS cut II LRGs. As sufficient sample 8 and 9 targets were not necessarily available in each 2SLAQ field a small number of `fibre-fillers' were selected below the Sample 9 cut. These are even lower redshift LRGs in the region sampled by SDSS cut I.

The 2SLAQ observations were performed using exclusively the 2dF spectrograph 2 with a 600 lines per mm $V$ grating. This gives a dispersion of 2.2\AA\/ per pixel and a resolution of about 5\AA\/. The detector used was a Tek1024 CCD with 1024 $\times$ 1024 pixels. Thus using 6150\AA\/ central wavelength the resulting spectra have a wavelength coverage from 5050\AA\/ to 7250 \AA\/. This coverage was chosen to ensure that the Ca H\&K lines were always present in the spectra across the target redshift range of the survey.

In this study we select 2SLAQ LRGs which fall into the Sample 8 and Sample 9 selections. In addition only LRGs which have confident redshifts are selected by requiring the redshift quality flag be greater than or equal to 3. Quality flags on 2SLAQ LRGs are set by the redshift code Zcode, a derivative of the code used to determine redshifts for the 2dFGRS survey. A quality flag of 3 or greater indicates a high level of confidence in the stated redshift (see Cannon et al. 2006 for details). 

We also require that both the \OII\/ and \hdel\/ lines are within the wavelength coverage of the spectra and that neither line is in close proximity to the strong sky emission lines at 5577\AA\/, 5900\AA\/, 6300\AA\/ and 7244\AA\/. The signal-to-noise ratio per pixel, averaged across the spectrum, is also required to exceed 3.5 as it was found that lower signal to noise data produces too many false line detections, as will be discussed in Section \ref{sec:def}

These requirements bring the number of 2SLAQ LRGs utilised in this study to 5,697. It should be noted that all magnitudes and colours throughout are on the ABmag scale, in line with the 5-band SDSS photometric system (see \nocite{fukugita1996}{Fukugita} {et~al.} 1996).

\section{Methodology} \label{sec:meth}
\subsection{Measuring Equivalent Widths}\label{sec:methew}
 
Equivalent widths (EW) of known spectral features are measured using a pseudo-Lick passband flux adding approach. The flux in a set band around the known line wavelength is computed and compared to a continuum level. The continuum is derived by calculating the average flux level in two sidebands and interpolating across the region of interest using these values. For the \hdel\/ and H$\gamma$ lines we use the Lick standard \hdel$_A$ and H$\gamma_A$ definition for the passband and sidebands \nocite{worthey1997} (Worthey \& Ottaviani 1997), while for the \OII\/ line we use the definition from Balogh \etal (1999). These definitions are presented in Table \ref{tab:mocklines}. It should be noted that the passband definitions for the absorption lines are the same as those used in the output from Bruzual and Charlot (2003) models. Note that throughout this paper we define a negative equivalent width as emission and a positive value as absorption in a stated line.

D4000 break strengths are also measured for each spectrum in the sample. We define the D4000 index as per \nocite{balogh99}{Balogh} {et~al.} (1999), namely the ratio of the flux in two 100\AA\/ windows centred on 4050 and 3900\AA. While the 2SLAQ spectra are not flux calibrated, we find that the change in spectrograph response is insignificant compared to the intrinsic error in the spectrum and thus the D4000 indices are unaffected.

Errors on the equivalent widths are calculated utilising equation A8 from \nocite{boh83}{Bohlin} {et~al.} (1983). Errors in the continuum are based on the standard error in the mean of the sideband flux; errors in the line are based on variance spectrum calculated by the 2dFDR pipeline. It should be noted that while the equivalent width, and D4000, errors take into account `random' errors such as photon counting statistics, they do not include contributions from systematic errors in the spectra, such as poor sky-subtraction or flat-fielding. Contributions from these sources to the equivalent width measure could be as high as a few percent, and as such the quoted errors are likely underestimates of the true error. While this may be of concern for future users of the data presented in Table 1, this is not considered a problem here as the error on the indices does not play a major role in any of the results presented.

A listing of the equivalent width measures on the \OII\/ and \hdel\/ line, as well as the D4000 indices for all \numl\/ 2SLAQ LRGs used in this study can be at \url{http://lrg.physics.uq.edu.au/publi.html}.

\section{Results} \label{sec:results}
\subsection{Definitions based on \OII and \hdel}\label{sec:def}
In this section we present a number of spectral classifications based on the \OII\/ and \hdel\/ equivalent widths which will be used in the subsequent analysis. We split the LRGs into 4 categories based on their position in the \EWH-\EWO\/ space. 

\begin{enumerate}
\item Passive - LRGs with no significant \OII\/ emission (\EWO$>-8$\AA) or \hdel\/ absorption (\EWH$<2$\AA), indicating an old passively evolving stellar population.
\item k+a - LRGs with significant \hdel\/ absorption (\EWH$>2$\AA) but no \OII\/ emission (\EWO$>-8$\AA).
\item em - LRGs with significant \OII\/ emission (\EWO$<-8$\AA) but no \hdel\/ absorption (\EWH$<2$\AA).
\item em+a - LRGs with significant \hdel\/ absorption (\EWH$>2$\AA) and significant \OII\/ emission (\EWO$>2$\AA).
\end{enumerate}

The implementation of this classification scheme quantitatively involves a careful determinination of what a `significant' amount of \OII\/ emission or \hdel\/ absorption is in light of the quality of the dataset. Physically the divisions in \EWH--\EWO\/ space should be designed in such a way to isolate LRGs which have star formation histories that are distinctly different from simply passive evolution. In the high signal-to-noise, and resolution limit this is relatively easy as passive evolution models have no \OII\/ emission and low levels of \hdel\/. However in the low signal-to-noise, low resolution regime, in which the 2SLAQ LRG spectra predominately lie, variations in the \EWH\/ and \EWO\/ will be dominated by noise rather than intrinsic physical properties. Thus rather than use a physical justification for our \EWH\/ and \EWO\/ divisions, we calculate threshold values at which it is unlikely that an LRG has a spectrum of a passively evolving stellar population in the presence of noise. 

To achieve this we introduce `mock' \hdel\/ and \OII\/ passbands into our equivalent width analysis. The wavelengths of these `mock' passbands are positioned in quiet regions of the LRG spectrum away from known spectral lines such that any significant equivalent width measures will be purely a result of noise. The passbands and sidebands used in the mock analysis are found in Table \ref{tab:mocklines}. For each spectrum in the 2SLAQ LRG sample the equivalent width of the `mock' line index is measured using the method described in Section \ref{sec:methew}. As well as giving us a benchmark with which to determine the optimum \EWH\/ and \EWO\/ threshold values, this analysis has the advantage that for any arbitrary set of \EWH\/ and \EWO\/ thresholds we have a good estimate for the false detection rate, i.e. the number of LRGs we expect to fall into our k+a, em+a or em sample due simply to noise. 

\begin{table*}

\caption{Passbands and side continuum bands used to measure the mock \hdel\/ and mock \OII\/ equivalent widths.}
\label{tab:mocklines}
\centerline{
\begin{tabular}{|l|l|l|l|}
\hline
 Index &Line passband (\AA)&Blue continuum sideband (\AA)&Red continuum sideband(\AA)\\
\hline
\hdel & 4083.5 -- 4122.25 & 4041.6 -- 4079.75 & 4128.5 -- 4161.0 \\
\OII & 3713 -- 3741 & 3653 -- 3713 & 3741 -- 3801 \\
 H$\gamma$ & 4319.75 -- 4363.5 & 4283.5 -- 4319.75 & 4367.25 -- 4419.75 \\
mock \hdel &4230.5 -- 4242.25& 4165.6 -- 4199.75& 4245.5 -- 4271.0\\
mock \OII &3590.0 -- 3617&3480.0 -- 3550.0&3617.0 -- 3641.0\\
\hline
\end{tabular}
}
\end{table*}

\begin{figure}
%\centering
%\begin{minipage}{140mm}
  \includegraphics[angle=270,scale=0.6]{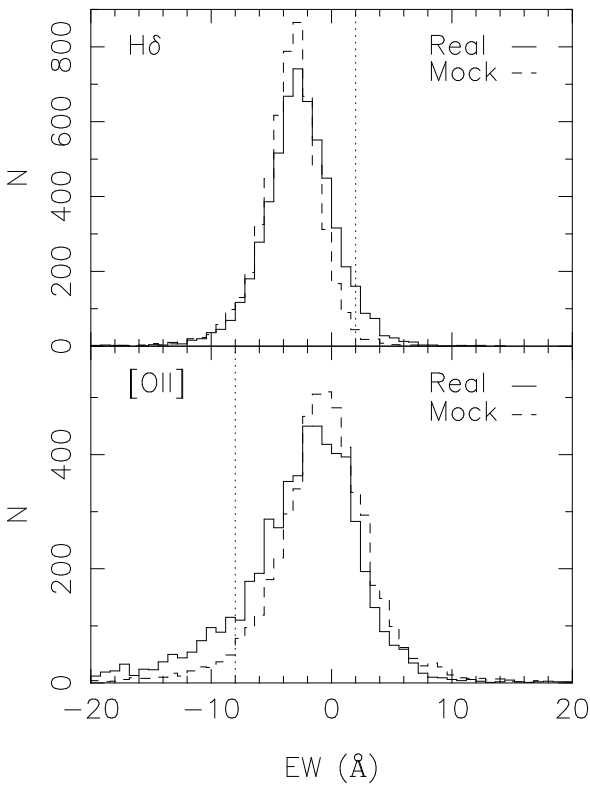}
  \caption{Distribution of real \hdel\/ (top) and \OII\/ (bottom) equivalent widths (solid line) and mock \hdel\/ (top) and \OII\/ (bottom) equivalent widths (dashed line). The low signal to noise of the 2SLAQ spectra is apparent from this figure as the mock equivalent width distributions cover a similar range to the real equivalent width distributions despite the fact that the there should be no strong features in the passband regions in the mock analysis. The dotted lines shows the equivalent width thresholds which define the classification scheme used throughout the paper.}
\label{fig:EWdist}
%\end{minipage}

\end{figure}

Figure \ref{fig:EWdist} shows the distribution of \OII\/ and \hdel\/ equivalent widths for both the real and mock index definitions. Note again that negative equivalent widths indicate emission while positive indicate absorption. The low signal to noise of the 2SLAQ data is apparent in the spread of the mock equivalent widths for both lines. The distribution of mock equivalent widths covers the same range as the real equivalent widths for both \hdel\/ and \OII\/, meaning it is impossible to define any divisions which will eliminate all false positives. Thus we try and define divisions that will minimise the number of false positives while still selecting a statistically significant number of LRGs in each of our classifications. It can be seen that the \hdel\/ mock equivalent width distribution drops significantly at $\sim 2$\AA, creating a large difference between the real and mock distributions. If we use this value to define `significant' \hdel\/ absorption the number of combined false positives in the k+a and em+a categories is 102, while the number of LRGs with real \EWH$>2$\AA\/ is 330. Thus using a division at \EWH=2\AA\/ results in 30.1 per cent of k+a and em+a's being false positives. If we decrease the division to 1\AA\/ the ratio of false positives to real detections increases to 33.5 per cent, while increasing the threshold to 5\AA\/ only decreases the ratio to 29.9 per cent, hardly worth a decrease in number of real detections of 275 (600 per cent).
  
By a similar justification we define the \EWO\/ division. The distribution of mock OII equivalent widths is seen to drop significantly at $\sim-8$\AA. Using this value as our \EWO\/ division gives a false positive to real detection ratio of 43.4 per cent. Decreasing to -5\AA\/ gives a ratio of 52.7 per cent, while increasing to -20\AA\/ gives a ratio of 54.9 per cent.

\subsection{Spectral Properties of 2SLAQ LRGs}\label{sec:rsf}

Using the selection criteria outlined in the preceding section, we can determine the fraction of LRGs in each of the four categories. Table \ref{tab:fract} shows the number of galaxies in each category, and what fraction of the total population they constitute.

\begin{table*}
\caption{Fraction of each spectral type and properties of their combined spectrum.}
\label{tab:fract}
\centerline{
\begin{tabular}{l|l|l|l|l|l|l|l}
\hline
Spectral Type & Number & Fraction (per cent) & Corrected fraction (per cent)& D4000 & \EWO (\AA) & \EWH (\AA) & EW$\{$H$\gamma\}$ (\AA)\\
\hline
Passive & 4612 & $81 \pm 1$ & $88\pm1$ &$1.75 \pm 0.01$& $-1.0 \pm 0.1$ & $-1.67 \pm 0.08$  & $-5.83 \pm 0.08$ \\
k+a & 225 & $3.9 \pm 0.3$ & $2.7\pm0.2$ & $1.53 \pm 0.01$ & $-2.3 \pm 0.1$ &$3.7 \pm 0.1$ &  $-1.7 \pm 0.01$ \\
em+a & 101 & $1.8 \pm 0.2$ & $1.2\pm0.1$ &$1.50 \pm 0.01$ &$-12.1 \pm 0.1$ & $3.9 \pm 0.1$ &  $-1.6 \pm 0.1$\\
em & 759 & $13.3 \pm 0.5$ & $ 8.6\pm0.4$& $1.71 \pm 0.01$ & $-11.7 \pm 0.1$ &$-1.37 \pm 0.09$ &  $-5.53 \pm 0.08$\\
\hline
\end{tabular}
}
\end{table*}

Errors on the stated fractions are calculated assuming Poisson statistics.  

The corrected fractions are calculated by subtracting the estimated false detection rate calculated in Section \ref{sec:def} from the measured fraction. The false detection rates for the \hdel\/ and \OII\/ line are transformed into rates for the k+a, em+a and em classifications by assuming that the false detection rate for a line is spread evenly over the classifications requiring it. For example, 30.1 per cent of \hdel\/ detections in k+a and em+a LRGs are false detections, thus if they are spread evenly across both classifications the false detection rate in k+a and em+a LRGs (due to only \hdel\/) is 20.8 and 9.3 per cent respectively. In the case of the em+a classification we define the false detection rate to be sum of the false detection rate from both \hdel\/ and \OII. 

The efficiency of the selection method in isolating passively evolving ellipticals is again apparent from Table \ref{tab:fract} with over 80 per cent of LRGs having the spectral features indicative of an old passively evolving stellar population.

%While only $2.7 \pm 0.02$ per cent of LRGs are classified as k+a, this fraction is reasonably consistent with that found in low redshift surveys. If we increase our \hdel\/ threshold to 5\AA\/, roughly matching that used in the \nocite{Zab96}{Zabludoff} {et~al.} (1996), \nocite{Blake2004}{Blake} {et~al.} (2004), and \nocite{goto2003b}{Goto} {et~al.} (2003) k+a studies, we find our k+a LRG fraction decreases to $0.43 \pm 0.09$ per cent, roughly comparable to the $0.2$, $0.15$ and $0.09$ per cent k+a fractions found in the those studies respectively. Thus while our colour selected sample is quite different to these studies we are seeing roughly the same proportion of `strong' k+a type galaxies as in purely magnitude limited samples at low redshift.

\begin{figure}

\includegraphics[angle=270,scale=0.7]{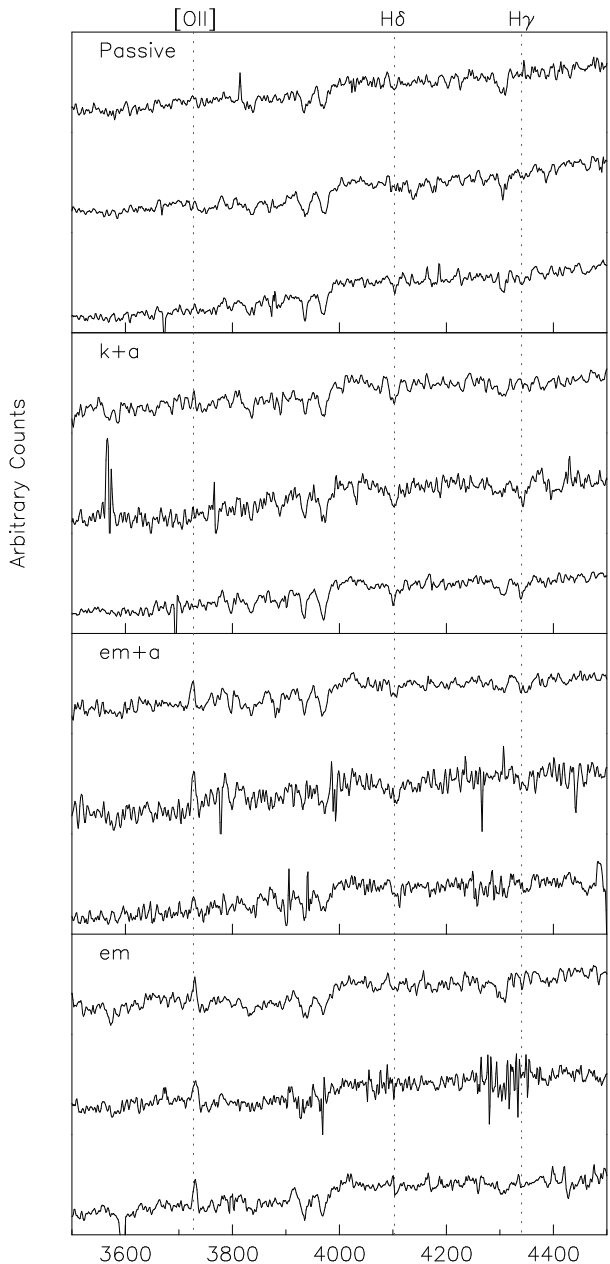}
\caption{Example spectra for the k+a (top three), em+a (middle three) and em (bottom three) spectral classifications. Spectra have been converted to rest frame wavelengths, and smoothed by a Gaussian with FWHM=0.8\AA\/. Spectral features discussed in the text are marked. }
\label{fig:exspec}

\end{figure}

A number of example spectra for each of the spectral classes are shown in Figure \ref{fig:exspec}. Spectral features discussed are marked. It can be seen that in the spectra with strong \hdel\/, the other, higher order, Balmer lines are also observed.

\begin{figure}

\includegraphics[angle=270,scale=0.6]{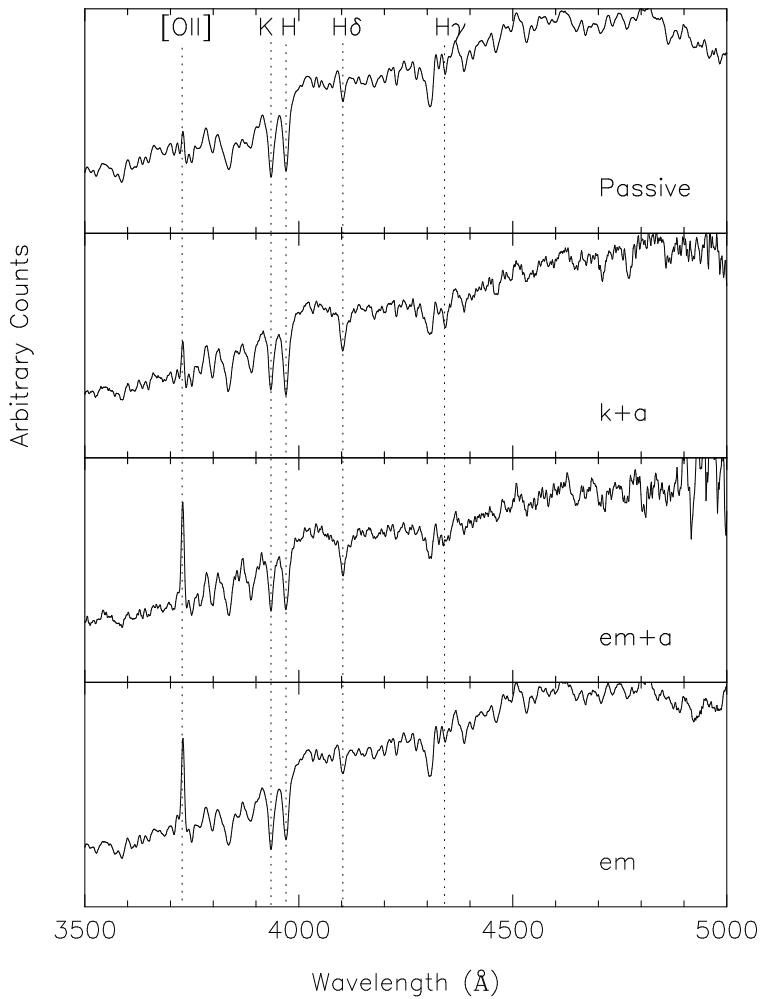}
\caption{Combined spectrum for the four spectral classifications. Quantities measured on the spectra are presented in Table \ref{tab:fract}. Note the increase in the absorption strength of the higher order Balmer lines (H$\gamma$,H$\epsilon$, etc) in the the k/em+a combined spectra.}
\label{fig:combspec}

\end{figure}

The combined spectra for each spectral type are presented in Figure \ref{fig:combspec}. The combined spectra are produced by first de-redshifting each spectrum, and rebinning to a common dispersion. The spectra are first normalised using the mean flux in a 100\AA\/ window (4000-4100\AA\/). The top and bottom 10 per cent of values at each pixel are then excluded. The mean of the remaining values at each pixel is then used to produce the combined spectrum. For each combined spectrum the D4000 index and equivalent widths of the \hdel\/, H$\gamma$, H$\beta$ and \OII\/ lines have been measured using the methods presented in Section \ref{sec:methew}, with the results presented in Table \ref{tab:fract}. 

While it is generally accepted that \hdel\/ is an efficient indicator of a young stellar population, the presence of strong \hdel\/ absorption should be accompanied by increased absorption in the other Balmer lines. To test this we measure the equivalent width of the H$\gamma$ line in the combined spectra, with the results presented in Table \ref{tab:fract}. H$\gamma$ alone is utilised as the lower order Balmer lines (H$\beta$, H$\alpha$) are not inside the wavelength coverage of many of the higher redshift spectra. It can be seen that for classifications which pass the \EWH$>2$\AA\/ cut, k+a and em+a, the equivalent width of H$\gamma$, is significantly greater in absorption than that found in the passive combined spectrum. It should be noted that the H$\gamma$ equivalent widths quoted in Table \ref{tab:fract} appear to be in emission (i.e. negative) as emission in the nearby passbands used to estimate the continuum dominates moderately low levels of absorption in the line itself. The ratio of the Calcium H to K lines is also noticably greater in the k+a and em+a combined spectrum than in the Passive spectrum (Figure 4), This can be attributed to increased absorption in the H$\epsilon$ line, which is coincident with the Calcium H line. Thus we can be satisfied that our selection based on \hdel\/ alone has been effective in isolating LRGs with overall increased Balmer line absorption, and hence potentially younger stellar populations.

\begin{figure}

\includegraphics[angle=270,scale=0.6]{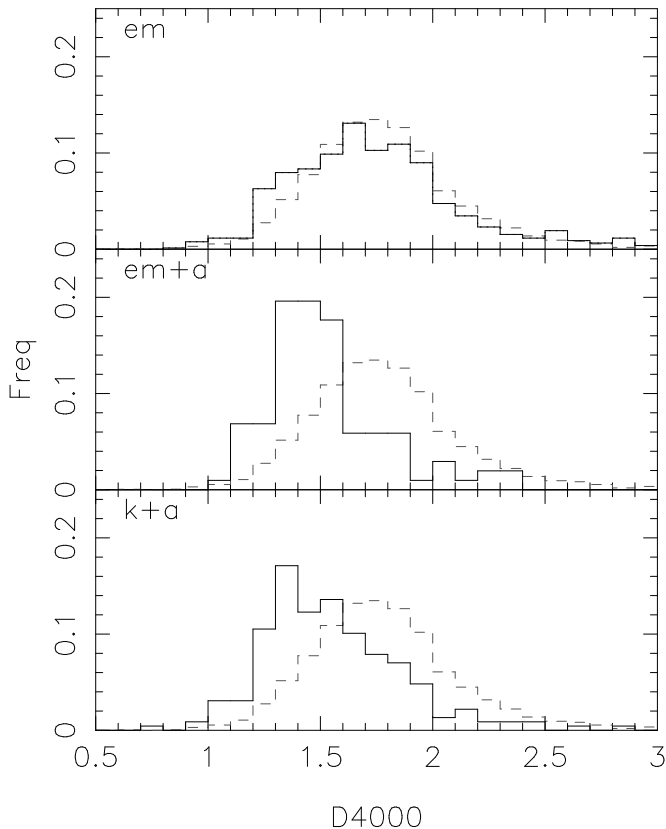}
\caption{Distribution of D4000 indices for k+a (bottom), em+a (middle) and em (top) LRGs. In each case the distribution is marked with a solid line, while the distribution of passive LRGs is marked with a dashed line. The k+a and em+a LRGs show systemically lower D4000 indices than the passive LRGs. This, along with their increased \hdel\/, confirms the presence of a younger stellar population in these LRGs. Significant numbered of em LRGs show D4000 strengths that are as large as those found in the passive population, suggesting that a large fraction of em LRGs do not possess significant populations of young stars.}
\label{fig:d4000hist}

\end{figure}

Figure \ref{fig:d4000hist} shows the D4000 distributions for the k+a (bottom), em+a (middle) and em (top) spectral classifications. The whole 2SLAQ sample is designated by the solid line, while the D4000 distribution of only the passive LRGs is shown by the dashed line. The D4000 index is well known to be sensitive to recent star formation, with lower values indicating the increasing presence of a young A star population. As expected the D4000 indices for LRGs in the 2SLAQ sample suspected to have recent star formation,(k+a \& em+a) are noticably less than that for the passively evolving LRGs. The em+a LRGs also show significantly lower D4000 indices than the k+a LRGs suggesting that the em+a galaxies have experienced more recent, or ongoing given the presence of the \OII\/ line, star formation. Interestingly the em LRGs have D4000 indices which are broadly consistent with the passive LRGs. While a KS test shows that all three distributions are statistically inconsistent with the passive population, with p-values of 2.4$\times10^{-6}$, 2.6$\times10^{-17}$ and 6.6$\times10^{-22}$ for the em, em+a and k+a LRGs respectively, a significant fraction of the em LRG population are found to have D4000 indices as large as those found in the passive population. This is in stark contrast to the k+a and em+a LRG populations which are very rarely found at high D4000 index (i.e. D4000$>2$). This would suggest that while recent star formation is unambigously responsible for k+a and em+a LRGs, the origins of em LRG population are not so clear.

\subsection{Redshift Dependence} \label{sec:zdep}
To test the redshift dependence of the classification fractions, the 2SLAQ LRGs are divided into 2 subsamples between $0.45<z<0.55$ and $0.55<z<0.65$. In order to increase the redshift range of the analysis a low redshift sample for comparison is provided by SDSS MAIN survey galaxies with $0.1<z<0.2$. While the SDSS also conducted an LRG survey the selection criteria for SDSS LRGs is much stricter than for 2SLAQ, so strict that only $\sim$700 2SLAQ LRGs would be selected by in the SDSS survey at $z=0.2$. It is for this reason that we utilise galaxies from the SDSS MAIN survey for comparison. 

\begin{figure}
\includegraphics[angle=270,scale=0.6]{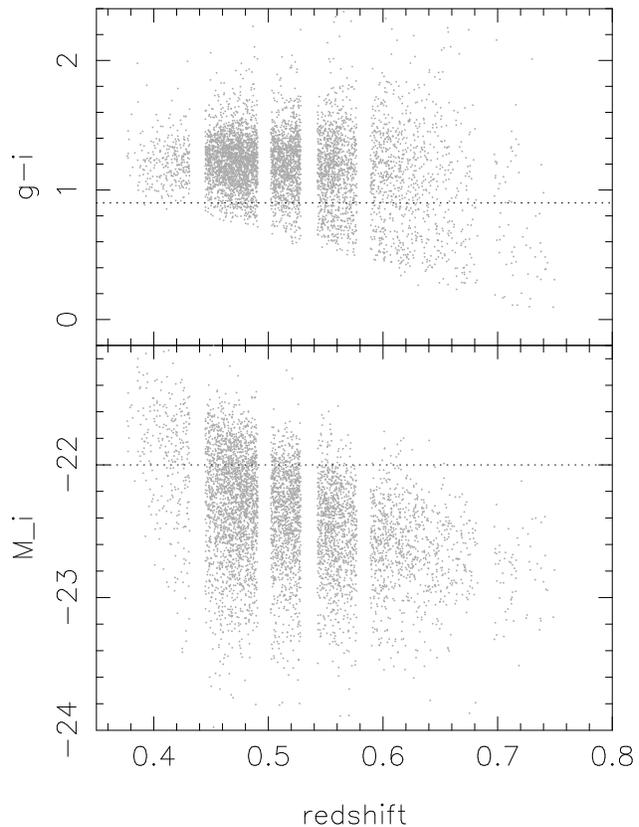}
\caption{Rest-frame $g-i$ and $M_i$ redshift evolution for the 2SLAQ LRGs. Rest-frame quantities are produced via k+e corrections based on a Bruzual and Charlot (2003) model track. The model assume a single epoch of star formation for the LRG stellar population at high redshift, followed by passive evolution. The dashed line shows the cuts utilised in the text to produce a sample of LRGs which are unbiased in colour and luminosity with redshift.} 
\label{fig:photevol}
\end{figure}

As the 2SLAQ LRGs were selected by an observed-frame colour selection some effort must be made to ensure our comparative sample from SDSS is consistent with the colour selections of 2SLAQ. Indeed effort must also be made to ensure that the colour selections are consistent within the redshift range covered by 2SLAQ itself. The simplest way to accomplish this is to use an independent rest frame selection in colour-magnitude space on both the 2SLAQ and SDSS MAIN galaxies. To determine the appropriate boundaries of this selection we present the M$_i$ and rest frame $g-i$ evolution of the 2SLAQ galaxies with redshift in Figure \ref{fig:photevol}. Absolute magnitudes for the 2SLAQ galaxies are produced via the use of model k+e corrections based on \nocite{Bruz03} Bruzual and Charlot (2003;henceforth BC03) models. The models assume simple passive evolution of the stellar population from a single epoch of high redshift star formation. This sort of model has been found to approximate the colour evolution of LRGs to $\sim0.1$ mag \nocite{wake2006} (Wake \etal 2006), which is satisfactory for our purposes here. While ideally k-corrections could be derived directly from the spectra, the 2SLAQ spectra are not flux-calibrated, nor do they have the significant wavelength coverage required for such calculations. Significant redshift evolution is shown in both the $g-i$ colour and M$_i$ of the LRGs in Figure \ref{fig:photevol}. The M$_i$ evolution is a result of the $i_{Dev}<19.8$ magnitude limit of the 2SLAQ survey, while the increase in the distribution of $g-i$ colours to bluer values is a result of the colour selection becoming `looser' at higher redshifts (as can be seen to some extent in Figure \ref{fig:ccpaper}. It can be seen that a cut at $g-i>0.8$ (as shown by the dashed line) will produce a sample which is reasonably consistent in colour with redshift. Similarly a cut at $M_i<-22$ will produce a sample which is close to volume-limited. It is on this cut-down `homogeneous' sample of 2SLAQ LRGs that the following analysis is performed.

In a similar fashion we select galaxies from the SDSS MAIN survey which have rest-frame $g-i>0.9$ and M$_i<-22.0$. As the SDSS spectra have a wide wavelength coverage, and good spectrophotometric calibration k corrections can be derived directly from the spectra. This has been done for the DR4 MAIN galaxies by \nocite{Blanton2005} Blanton \etal\/ (2005) and can be found in the NYU SDSS value-added catalogue (VAC) available at \url{http://wassup.physics.nyu.edu/vagc/}. K corrections in the NYU VAC are produced via the use of the kcorrect software (v3.4) described in \nocite{Blanton2003} Blanton \etal\/ (2003). As the VAC only provides k-corrected photometry the M$_i$ cut is adjusted slighty to -22.2 to take into account of the M$_i$ evolution from $z=0$ to $z=0.15$ predicted by the BC03 models used to k+e correct the 2SLAQ LRGs.

\begin{figure}
\includegraphics[angle=270,scale=0.35]{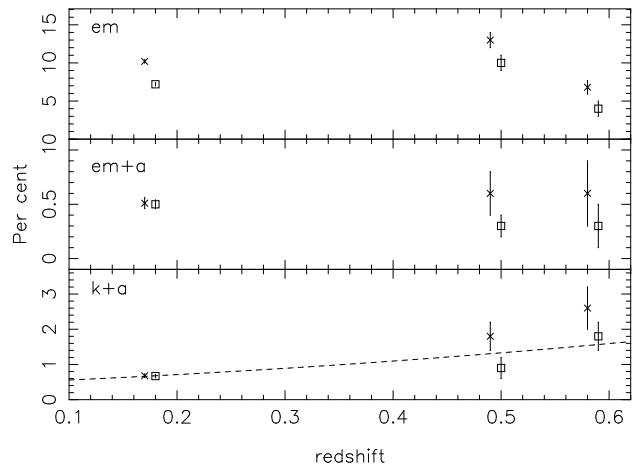}
\caption{Redshift dependence for the raw (crosses) and corrected (squares) fraction of each classification with redshift. For comparison the fraction of each classification in a sample of low redshift ($z\sim 0.15$) SDSS MAIN galaxies is shown. The SDSS MAIN galaxies are required to have the same luminosity and colour range as the 2SLAQ LRGs by placing simple cuts in rest-frame $g-i$ and $M_i$. While the em and em+a fractions do not show any overall trend with redshift, the k+a fraction shows a strong increase. Fitting a $(1+z)^n$ relation to the k+a fraction increases gives $n=2.8\pm0.7$. The fit is shown by the dashed line.}
\label{fig:corrfrac}
\end{figure}

The results of this analysis are presented in Figure \ref{fig:corrfrac} and Table \ref{tab:fracevol}. In each subsample from 2SLAQ, and for the low-$z$ SDSS galaxies, we calculate the number of each spectral classification present, and the fraction of the bin that this number constitutes. Errors are again calculated assuming Poisson statistics. Since the LRGs are not uniformly distributed inside each bin the bin centres are taken to be the median redshift of the all the LRGs in a given bin.

\begin{table*}
%\centerline{
\caption{Redshift dependence of classification fractions. Quoted errors are calculated assuming Poisson statistics. The fractions are calculated by first producing a truncated sample which is uniform in colour and luminosity with redshift, and then correcting the fractions using the same analysis as described in Section \ref{sec:def}. The low redshift $z\sim0.17$ point is measured using SDSS MAIN galaxies with the same rest-frame colours and luminosities as the 2SLAQ LRGs.}
\label{tab:fracevol}
\begin{tabular}{l|l|l|l|l|l|l|l|l|l|l}
\hline
$z$ & N[LRGs] & N[k+a] & Per cent  & Corrected & N[em] & Per cent & Corrected & N[em+a] & Per cent & Corrected \\
\hline
0.17 & 20781 & 143& $0.68\pm 0.05$& $0.67 \pm 0.05$ &2115 & $10.2\pm 0.2$ &$7.2\pm0.2$ &107 & $0.51 \pm 0.04$ & $0.50\pm0.04$\\
0.49 & 2610 & 46 & $1.8\pm 0.4$&$0.9\pm0.3$ &303 &$13\pm1$&$10\pm1$ &16&$0.6\pm0.2$& $0.3\pm0.1$\\
0.57 & 1141 & 33 & $2.9\pm0.6$&$1.8\pm0.4$ &78&$6.8\pm0.9$&$4\pm1$ &7&$0.6\pm0.3$& $0.3\pm0.2$\\
\hline
\end{tabular}

%}
\end{table*}

In order to correct for the occurrence of false detections we again apply corrections to the fractions estimated using the results of the `mock' line analysis from Section \ref{sec:def}. By dividing the `mock' equivalent width measures on \hdel\/ and \OII\/ into the same redshift bins as the real analysis we produce corrections which are indicative of the underlying signal to noise distribution in each bin. This has the advantage of also eliminating any possible bias that may arise from varying signal to noise across the redshift range.

The raw and corrected fractions are presented in Figure \ref{fig:corrfrac} with the values shown in Table \ref{tab:fracevol}.
 
The effect of the $g-i>0.9$ and M$_i<-22$ cuts on the number of k+a and em+a LRGs is quite dramatic. Only 79 out of the original 225 k+a's and even more dramatically only 23 out of the original 101 em+a LRGs pass the cuts. However this is not all that surprising given the strong correspondence between k+a and em+a features and low D4000, which can be taken as an effective proxy for $g-i$ colour. 

It can be seen from Figure \ref{fig:corrfrac} that only the k+a LRG fraction shows a clear trend of increasing with redshift. The em LRG fraction shows a strong increase between the SDSS $z=0.17$ sample and 2SLAQ LRGs at $z=0.49$, but then shows a significant drop in the highest redshift bin. Little can be said about the em+a LRG fraction as the numbers in the 2SLAQ bins are too low. 

In order to quantify the redshift dependence of the k+a fraction in Figure \ref{fig:corrfrac} a $(1+z)^n$ fit has been performed on the corrected fractions. Fitting is performed via weighted linear regression in log$(z)$ vs log(fraction) space. The fit is shown by the dashed line in Figure \ref{fig:corrfrac}. The k+a fraction is found to obey a $(1+z)^{2.8\pm0.7}$ relation. 

\section{Discussion}\label{sec:disc}

\subsection{Comparison with previous work}
In recent years many authors have looked for recent and/or ongoing star formation amongst samples of early-type galaxies. \nocite{Fukugita2004} Fukugita \etal\/ (2004) performed a spectroscopic study of 420 E/S0 $z\sim 0.12$ galaxies selected from SDSS. They found evidence of ongoing star formation, in the form of H$\alpha$ emission, in 19 (4.5 per cent) and k+a like spectral properties in 15 (3.6 per cent). \nocite{Yi2005} Yi \etal\/ (2005) also identified early-types with recent star formation in SDSS, this time using GALEX Near-UV and SDSS $r$-band photometry. $\sim15$ per cent of their sample of 39 bright (M$_r<-22$), early-type, $z<0.13$ galaxies was found to have evidence of recent star formation via the use of NUV-$r$ colour. Moving to higher redshifts Doherty \etal\/ (2005) and Le Borgne \etal\/ (2005) both find significant fractions ($\sim 30$, and 50 per cent, respectively) of k+a's amongst massive galaxies at $z\sim1$.

A significant fraction of 2SLAQ LRGs fall into the em category ($6.8\pm 0.3$ per cent), although this result is not completely unexpected as a study by \nocite{Willis2002}{Willis} {et~al.} (2002) found over 20 per cent of their sample of luminous field early type galaxies had significant \OII\/ emission. Similarly \nocite{eis03} Eisenstein \etal\/ (2003) found that $\sim$ 10 per cent of their sample of massive bulge dominated galaxies had emission line components.

However it is difficult to make concrete comparisons between these results and those presented here. This is because the fraction of early-types with young stellar populations is very sensitive to the sample selection. This can even be seen here, with significant differences in the fraction of k+a and em+a LRGs between the whole sample in Section \ref{sec:rsf} and cut-down sample used in Section \ref{sec:zdep}. Most previous studies, including those discussed above, use a selection based on morphological information. This is distinctly different from the colour based selection used in 2SLAQ survey. It should be expected that morphology based studies of early-types will find greater fractions of galaxies with young stellar populations than we find here as the colour selection is strongly biased against the selection of galaxies with large populations of young stars. Thus while the absolute fractions of k+a and em+a LRGs quoted here are not directly comparable to other work on massive early-types, the observed increase with redshift should be consistent between the studies.

This notion is not only supported qualitatively by the jump in the number of `young' early-types between $z\sim0.1$ and $z\sim1$ but also quantitatively by a comparison of the measured redshift dependence of the k+a fraction here and in the study of \nocite{leborgne2005} Le Borgne \etal (2005). Le Borgne \etal\/ (2005) performed a study of the spectral properties of massive (M$>10^{10.02}$M$_{\sun}$) galaxies selected from the Gemini Deep Deep Survey (GDDS; see \nocite{abraham2004} Abraham \etal 2004) and SDSS MAIN galaxy sample. While they did not base their selection on morphology or colour, and hence found a significantly greater number of k+a than we find, they found an evolution of (1+z)$^{2.5\pm0.7}$ across a redshift range of the $0.1<z<1.2$. Le Borgne \etal (2005) note that this value is very close to the predicted evolution of the galaxy merger rate in $\Lambda$CDM semi-analytic models of $(1+z)^{3.2}$ \nocite{lefevre2000}(Le F{\` e}vre \etal 2000). Interestingly the fraction of k+a and em+a LRGs found here is very close to $4\pm1$ per cent fraction of massive galaxies in close pairs found by Bell \etal\/ (2006). Thus if k+a and/or em+a activity is triggered by mergers, the fractions and evolutionary trends we find here are reasonably consistent with other merger indicators in the literature.

\subsection{\hdel\/ strong LRGs} \label{sec:hdelstrong}
% DEAL WITH THIS!!!!!!!!
If we accept the interpretation that the observed \hdel\/ absorption in the k+a and em+a LRGs is a result of recent star formation, which seems likely given the weaker D4000 indices and observed absorption in the higher order Balmer lines (see Section \ref{sec:rsf}), then a small, but significant, number of the LRG population ($\sim1$ per cent) has been forming stars within the last 2 Gyr. This is an important result as it shows that evolution is occurring in the massive early-type population at redshifts less than 0.7.   

To try and quantify the level of star formation required to produce the observed \EWH\/ we turn to the spectral synthesis models of \nocite{Bruz03}{Bruzual} \& {Charlot} (2003). A distinct advantage of using these models is that they produce output measurements of the indices discussed here (D4000 and \hdel\/) measured in same fashion as outlined in Section \ref{sec:meth}. All of the models presented in this Section assume a Salpeter (1955) IMF, solar metallicity, and no dust reddening. 

\begin{figure}
\centering
\includegraphics[angle=270,scale=0.35]{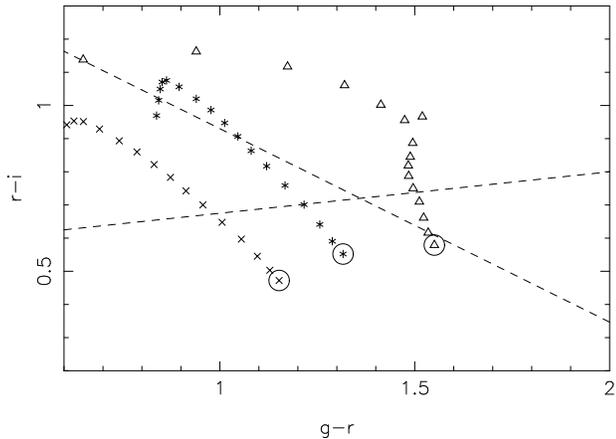}
\caption{Model $g-r$, $r-i$ tracks for two constant star formation models (crosses and asterisks), and a passive evolution model (triangles). The constant star formation models assume most of their stars are formed at high redshift, with a low level of constant star formation occurring since. The crosses represent a model in which 90 percent of the stellar population at age 10 Gyr was formed at high redshift, the asterisks represent a model where 95 per cent were formed at high redshift. The dashed line shows the Sample 9 2SLAQ selection described in Section \ref{sec:sam}. The passive evolution model assumes that all of the stars are formed at high redshift, and have experienced only passive evolution since. The circles indicate redshift of $z=0.4$ and the tracks mark the evolution from z=0.8 (near the upper left border) to $z=0.4$ at $\Delta z=0.025$ intervals. Assuming a formation age of $z=3$ for the model, the typical \EWH\/ in the redshift range probed by 2SLAQ is $\sim2.9$\AA\/ and $\sim1.6$\AA\/ for the 10 per cent and 5 per cent model respectively. Thus it can be seen that any galaxy experiencing a level of constant star formation high enough to pass our \EWH$>2$\AA\/ cut would not be selected by the 2SLAQ colour selection.}
\label{fig:consfmod}
\end{figure}

While constant star formation in a galaxy will produce the required level of \hdel\/, this scenario has two major problems. Firstly k+a LRGs show no \OII\/ emission, suggesting that there is no ongoing star formation occurring, however it is possible that the \OII\/ could be missing due to preferential dust obscuration or even aperture effects. Secondly, models show that galaxies with continuous star formation will mostly be outside of the LRG colour selection. Figure \ref{fig:consfmod} shows model tracks for two constant star formation models and a purely passive evolution model. In each of the constant star formation models a large fraction of the stellar population is formed at high redshift ($z\sim3$), with low levels of constant star formation responsible for 10 and 5 per cent of the stars at age 10 Gyr. The 10 per cent model track falls well outside of the 2SLAQ colour selection, while the 5 per cent model track just enters the 2SLAQ selection at $z>0.55$. The typical \EWH\/ of the model spectra in the redshift range probed by 2SLAQ is $\sim2.9$\AA\/ and $\sim1.6$\AA\/ for the 10 per cent and 5 per cent model respectively. Thus the 10 per cent model spectra would be selected in our k+a sample, but not in 2SLAQ and the 5 per cent model would be selected in 2SLAQ, but not as a k+a. As the 5 per cent model only just falls into the 2SLAQ selection we consider it unlikely that continuous star forming models can explain the k+a LRGs in 2SLAQ.

\begin{figure}
\centering
\includegraphics[angle=270,scale=0.35]{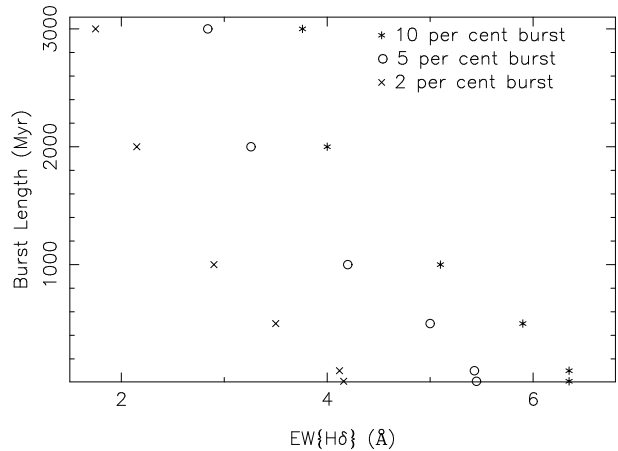}
\caption{Effect of changing the length of the star burst on the peak \EWH\/. Bursts consuming 2 (crosses), 5 (circles), and 10 (asterisks) per cent of galactic stellar mass are shown. The peak \EWH\/ is seen to decrease with increasing the burst length. It is likely that the star formation rate, not the total size of the burst, is the dominating factor in determining the peak \EWH\/. }
\label{fig:bursthd}
\end{figure}

Thus we confine ourselves to models in which a significant episode of star formation has occurred and subsequently ceased. In order to simplify the models we only consider scenarios in which an instantaneous burst of star formation is superimposed on an old stellar population. While it is likely that star formation histories with longer periods of star formation are responsible for the k/em+a population, the tracks taken by such models after star formation has ceased are nearly indistinguishable from the instantaneous burst models. The only quantitative difference is that longer star burst models require that a larger fraction of the LRGs mass be consumed in star formation to achieve the same level of \hdel\/ absorption. This can be seen from Figure \ref{fig:bursthd} which shows the relationship between burst length, and peak \EWH\/ for models which consume 2, 5, and 10 per cent of galactic stellar mass in new star formation. In all cases the models superimpose a burst of constant star formation on an old (7.4 Gyr) stellar population. The peak \EWH\/ is found to not only be dependent on the size of the burst (i.e. the total number of new stars created) but is also strongly dependent on the period over which the burst occurs. Indeed for bursts in the range 500-2000 Myr the peak \EWH\/ is dominated by the star formation rate, not the total size of the burst. This can be seen if we take an arbitrary peak \EWH\/ and look at the star formation rate (in per cent of galactic stellar mass per year) required for the three burst sizes presented in Figure \ref{fig:bursthd}. Using \EWH\/=4\AA\/ as an example, we estimate star formation rates for the 2 per cent, 5 per cent, and 10 per cent models of $6.6\times10^{-9}$, $4.2\times10^{-9}$ and $5.0\times10^{-9}$ per cent per year, respectively. It can be seen that these values are in relatively good agreement.

\begin{figure}
\includegraphics[angle=270,scale=0.45]{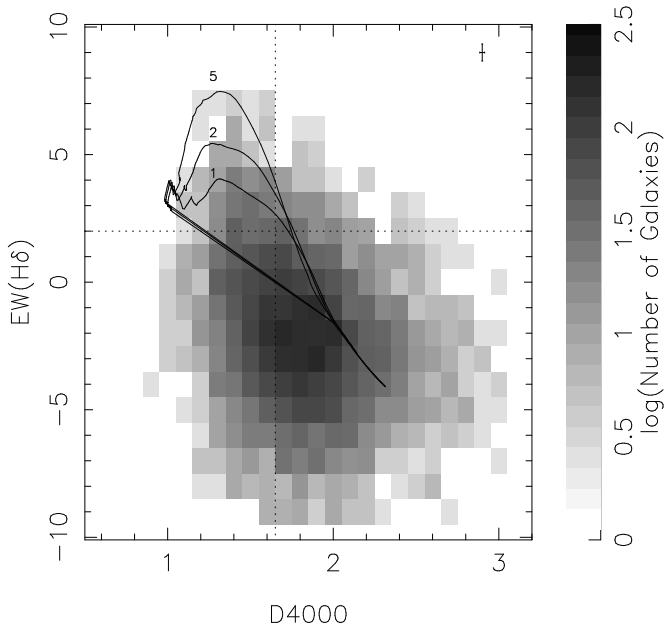}
\caption{Number density of LRGs in the D4000-\EWH\/ plane. The plane is broken down into 0.1x1 cells, the grey scale representing the logarithm of the number of LRGs in the cell. The scale goes from light to dark, with the darkest spot equal to 316 (10$^{2.5}$) LRGs. Also plotted are model tracks for a series of models assuming bursts of star formations consuming 5 (top), 2, and 1 (bottom) per cent of stellar mass in new star formation. Typical errors on both indices are shown in the top right corner. A significant number of LRGs have \EWH\/ and large D4000 indices which cannot be explained by these simple models. These are analogous to red \hdel-strong galaxies found in the literature (e.g. {Couch} \& {Sharples} 1987). }
\label{fig:mod4000}
\end{figure}

Figure \ref{fig:mod4000} plots the track taken in \EWH\/- D4000 space by models in which star bursts consuming 1, 2, and 5 per cent of stellar mass have occurred. In each case the model assumes an instantaneous burst occurs in a passively evolving stellar population which formed 7.4 Gyr before burst. Also shown in Figure \ref{fig:mod4000} is the number density of LRGs plotted in grey scale. As expected a slight trend of increasing \EWH\/ with decreasing D4000 is observed. Reasonable agreement is found between the \hdel\/ strong LRGs and the model tracks. Given that we have only presented tracks for a limited range of models it should be possible to produce models in a similar vein to those presented which can match any observed \EWH\/. However this is not the case for all values of D4000. The dotted vertical line at D4000=1.65 is used to crudely divide between those \hdel\/ strong LRGs which roughly agree with the models and those which don't. Of the 326 \hdel\/ strong LRGs in 2SLAQ 98 (30.1 per cent) have D4000$>1.65$. Many of the previous E+A/\hdel\/-strong studies have also identified a population of red \hdel\/ strong galaxies with photometric properties that are too red to be consistent with models \nocite{cou87,Blake2004,Pogg99, Balogh2005}({Couch} \& {Sharples} 1987; {Blake} {et~al.} 2004; {Poggianti} {et~al.} 1999; {Balogh} {et~al.} 2005). While dust and super solar metallicities can be called upon to redden the model tracks, and hence explain the discrepancies observed, it has been repeatedly shown that these effects, when used in plausible amounts, can only move the tracks shown in Figure \ref{fig:mod4000} by at most 0.1 in D4000 \nocite{balogh99,Pracy2005} (Balogh et~al. 1999, Pracy et~al. 2005), not nearly enough to explain the reddest \hdel\/ strong LRGs.  Curiously the fraction of red \hdel\/ strong LRGs in 2SLAQ is exactly equal to the predicted number of false positives in the sample. Given this it would seem probable that most, if not all, of the red \hdel\/ strong LRGs in this study are simply false detections. If this is true than the origin of the k+a and em+a LRG population can be most readily explained by recent bursts of star formation, on the scale of 1 per cent of galactic stellar mass, in otherwise passive LRGs.

\subsection{em LRGs}\label{sec:discem}

The origin of the em LRGs is more ambiguous as the observed \OII\/ emission could originate from either ongoing star formation or AGN. Typically the nature of emission lines in optical spectra can be determined via the use of line ratios ( e.g.\nocite{Baldwin1981} Baldwin et al. 1981), or even more unambigously by their radio properties. Unfortunately neither of these diagnositics are readily applicable here. The traditional optical line ratios used for classifying AGN/SF are H$\beta$, [OIII] 5007\AA, and H$\alpha$, all of which are have wavelengths too red to appear inside the wavelength coverage of 2SLAQ spectra. Sadler \etal\/ (2006) have performed a crossmatching of the 2SLAQ LRGs to radio sources in both the FIRST and NVSS catalogues. Of the 13784 2SLAQ LRGs with confident redshifts only 378 have corresponding radio sources in FIRST, this is despite FIRST covering 96.5 per cent of the 2SLAQ survey area. Of the 759 em LRGs found in the sample presented here, only 21 (2.8 per cent) have a FIRST radio detection within 5'', very close to the overall radio detection rate of 2.7 per cent. Similarly Sadler \etal\/ (2006) find that the fraction of radio detections with \OII\/ emission is the same as the fraction in the whole 2SLAQ sample. However while strong radio continuum emission can unambigously identify an AGN, the converse, i.e. lack of radio emission constitutes a lack of AGN activity, is not true. Thus while we can be confident that at least 2.8 per cent of em LRGs have AGN, little can be said about the remaining 97.2 per cent from available radio data.

However despite the lack of data with which to accurately classify each em LRGs, we can muse about the origins of em LRGs. A strong piece of evidence against a star forming origin for all em LRGs is the lack of correlation between \OII\/ emission and lower D4000 indices. The D4000 index is very sensitive to any population of young stars, as seen in the case of the k+a and em+a LRGs in Figure \ref{fig:d4000hist}. Thus unless we are observing an LRG during the very short period of time at the very beginning of a star formation episode before substantial numbers of young stars have been formed, their should be a clear relation between the D4000 index and current star formation. However if the \OII\/ emission originates from AGN activity this should have no discernable effect on the D4000 index. Given this it is unlikely that the large number of em LRGs with relatively large D4000 indices have star-forming origins. However given the large spread in the D4000 index seen in the passive LRG population (Figure \ref{fig:d4000hist}) it is impossible to say whether the em LRGs with low D4000 index are actively star forming. 

\subsection{Recent Star Formation in LRGs and the $\Lambda$CDM Paradigm} \label{sec:lcdm}
Assuming that recent and or ongoing star formation is the cause of the observed spectral characteristics of the k/em+a LRGs, which is likely given the arguments of the previous sections, a number of well-known phenomena can be called upon as potential candidates for origin of star formation in LRGs.

Given that LRGs are akin to massive early-type galaxies a sizable fraction will reside in groups or clusters, making a small, but significant, number of them bright cluster or even cD galaxies. Thus it is plausible that these central cluster LRGs may be at the focus of a cooling flow. The presence of optical emission lines in 20--30\% of all brightest cluster galaxies in X-ray selected samples has been confirmed in a number of studies (Donahue et al. 1992; Crawford et al. 1999) and an extreme population of galaxies with massive starbursts (50--200 M$_\odot$yr$^{-1}$) has been identified \nocite{allen99} (Allen 1995; Crawford et al. 1999). These galaxies could be the progenitors of the k/em+a LRGs if the starburst is truncated although they would be preferentially found in the brightest LRGs.

Another possibility is that the bursts of star formation are being created as a result of LRG-galaxy mergers or interactions. This ties in well with the idea that LRG evolution is dominated by hierarchical merging. Indeed in this model one would naively expect (and indeed it has been shown from N body simulations) that the frequency of mergers would decrease significantly with increasing time (decreasing redshift), which is qualitatively in line with the results demonstrated here. 

However it is difficult to make quantitative assertions about the origin of k/em+a LRGs as it is impossible to determine which of these hypotheses is responsible, if at all. Another complication is the potential for differences between the k+a and em+a classes. Previous studies of traditional k+a's and em+a analogues at low redshift have found systematic differences between the two classes, suggesting different formation mechanisms (Balogh \etal\/ 2005).

% fix this up!
Regardless of the physical nature of the star formation in LRGs, it is reasonably safe to assume that the star forming gas is coming from an external origin. This is in agreement with the core principles of $\Lambda$CDM cosmology. Results from the current generation of hydrodynamic N-body cosmological simulations suggest that the most massive galaxies increase in mass significantly from $z<1$ \nocite{delucia2005} (De Lucia \etal\/ 2005), with the bulk of this increase coming from accretion of intra-cluster gas, or via mergers with smaller galaxies \nocite{murali2002} ({Murali} et~al. 2002). By consideration of the expected timescale of the k+a signature we can crudely estimate the number of LRGs affected by k+a activity since $z=0.8$. Converting the redshifts of the bins used in Section \ref{sec:zdep} into look-back time results in a $0.36\times t^{0.85\pm0.2}$ dependence, where $t$ is the look-back time in Gyr. By simply integrating we can use this dependence to estimate the fraction of LRGs which undergo a k+a phase between $0<z<0.8$ (or $0<t<7$) as $8$ per cent. For comparative purposes we consider both the models of De Lucia \etal\/ (2005) and the observational evidence presented by Bell \etal\/ (2006). De Lucia \etal\/ (2005) predict that $\sim50$ per cent of massive early-types gain 50 per cent of their stellar mass between $0<z<0.8$. Similarly Bell \etal\/ (2006) predict from there measure of the close pair fraction that $\sim20$ per cent of massive galaxies will undergo a major merger between $0<z<0.8$. Thus even if k+a, or em+a, activity was associated with a major merger event that doubled the mass of the progenitor LRG, which is unlikely given the models presented in Section 5.2, this would only account for $\sim 16$ per cent of the required mass growth in massive early-types in the models, and only $40$ per cent of the predicted number of mergers in massive galaxies from observations. It is clear from this crude comparison that the level of k+a and em+a activity is not sufficient to explain the bulk of the mass growth in massive early types. This is in line with recent results which suggest that red or `dry' mergers are responsible for the growth of massive early-type galaxies (van Dokkum et~al. 2005; Bell et~al. 2005).

\section{Conclusion}\label{sec:conc}

We have determined the recent star formation histories of a sample of \numl\/ LRGs based on the equivalent widths of the \hdel\/ and \OII\/ lines. While the majority ($>$80 per cent) show the spectral properties of an old, passively evolving, stellar population, a significant number of LRGs show evidence for recent and/or ongoing star formation in the form of k+a (2.7 per cent), em+a (1.2 per cent) or em LRGs (8.6 per cent). By dividing the sample into 2 redshift subsamples from $0.45<z<0.55$ and $0.55<z<0.65$, and comparing to a $z\sim0.15$ sample selected from SDSS, it is observed that the fraction of k+a LRGs increases with redshift as $(1+z)^{2.8\pm0.7}$. 

Spectral synthesis models, utilising the code of \nocite{Bruz03}{Bruzual} \& {Charlot} (2003), suggest that the k/em+a LRGs could originate from passive LRGs which undergo a starburst in which material equivalent to $\sim 1$ per cent of the stellar mass is consumed in new star formation. 

Several origins for k+a and em+a LRGs are considered, including cooling flows and mergers, however identification of the exact formation mechanism, or even if k+a and em+a LRGs share a common origin, is not possible with the current data. It is clear that k+a and em+a activity represents a signpost of recent evolution in LRGs, however by considering the life time of the k+a/em+a phase we estimate that only 8 per cent of LRGs will experience a k+a or em+a phase between $z=0.8$ and the present. By comparing this with the semi-analytic galaxy formation models of De Lucia et~al. (2005) we conclude that k+a/em+a activity cannot be responsible for the predicted mass growth of massive ellipticals since $z=0.8$.

\section*{Acknowledgments}
IGR is supported by a UQCS scholarship. KAP is supported by an EPSA University of Queensland Research Fellowship and a UQRSF grant. ACE and IRS acknowledge support from the Royal Society.\\
The 2SLAQ survey was made possible through the dedicated efforts of the staff at the Anglo-Australian Observatory, both in creating the 2dF instrument and supporting it on the telescope. \\
Funding for the creation and distribution of the SDSS Archive has been provided by the Alfred P. Sloan Foundation, the Participating Institutions, the National Aeronautics and Space Administration, the National Science Foundation, the U.S. Department of Energy, the Japanese Monbukagakusho, and the Max Planck Society. The SDSS Web site is http://www.sdss.org/.\\
The SDSS is managed by the Astrophysical Research Consortium (ARC) for the Participating Institutions. The Participating Institutions are The University of Chicago, Fermilab, the Institute for Advanced Study, the Japan Participation Group, The Johns Hopkins University, the Korean Scientist Group, Los Alamos National Laboratory, the Max-Planck-Institute for Astronomy (MPIA), the Max-Planck-Institute for Astrophysics (MPA), New Mexico State University, University of Pittsburgh, University of Portsmouth, Princeton University, the United States Naval Observatory, and the University of Washington.\\

We also thank the anonymous referees for the many useful comments which greatly improved the paper.

\bibliography{}
 \bsp

\label{lastpage}

\end{document}